# Resistance Calculation for an infinite Simple Cubic Lattice Application of Green's Function

**J. H. Asad**[* †], **R. S. Hijjawi**[††], **A. Sakaj**[†††] **and J. M. Khalifeh**[† *]
[†]*Department of Physics, University of Jordan, Amman-11942, Jordan.*
[††] *Physics Department, Mutah University, Jordan.*
[†††]*Physics Department, Ajman University, UAE.*

Abstract

It is shown that the resistance between the origin and any lattice point $(l, m, n)$ in an infinite perfect Simple Cubic (SC) is expressible rationally in terms of the known value of $G_o(0,0,0)$. The resistance between arbitrary sites in a SC is also studied and calculated when one of the resistors is removed from the perfect lattice. Finally, the asymptotic behavior of the resistance for both the perfect and perturbed SC is also investigated.

- Corresponding author EMAIL: **jhasad1@yahoo.com**.

## 1. Introduction

The calculation of the resistance between two arbitrary grid points of infinite networks of resistors is a new-old subject[1–7]. Recently, Cserti[8] and Cserti et. al[9] studied the problem where they introduced for the first time a method based on the Lattice Green's Function (LGF) which is an alternative approach to using the superposition of current distributions presented by Venezian[3] and Atkinson et. al[4].
The LGF for cubic lattices has been investigated by many authors[10–20], and the so-called recurrence formulae which are often used to calculate the LGF of the SC at different sites are presented[21,22].
The values of the LGF for the SC have been recently evaluated exactly[23], where these values are expressed in terms of the known value of the LGF at the origin.
In this paper; we calculate the resistance between two arbitrary points in a perfect and perturbed (i.e. a bond is removed) infinite SC using the method presented by Cserti[8] and Cserti et. al[9].
The LGF presented here is related to the LGF of the Tight-Binding Hamiltonian (TBH)[24].

## 2. Perfect SC Lattice

In this section we express the resistance in an infinite SC network of identical resistors between the origin and any lattice site $(l,m,n)$ rationally, where it can be easily shown that [8,23]

$$\frac{R_0(l,m,n)}{R} = \rho_1 g_0 + \frac{\rho_2}{\pi^2 g_0} + \rho_3 \qquad (1)$$

where $g_0 = G_0(0,0,0)$ is the LGF at the origin.
and $\rho_1, \rho_2, \rho_3$ are related to $r_1, r_2, r_3$ (i.e $\lambda_1, \lambda_2, \lambda_3$ Duffin and Shelly's parameters[23,25]) as

$$\rho_1 = 1 - r_1 = 1 - \lambda_1 - \frac{15}{12}\lambda_2$$

$$\rho_2 = -r_2 = \frac{1}{2}\lambda_2 \qquad (2)$$

$$\rho_3 = -r_3 = \frac{1}{3}\lambda_3$$

Various values of $r_1, r_2, r_3$ are shown in Glasser et. al[23] [Table 1] for $(l,m,n)$ ranging from $(0,0,0)-(5,5,5)$. To obtain other values of $r_1, r_2, r_3$ one has to use the following relation[22]

$$G_0(l+1,m,n)+G_0(l-1,m,n)+G_0(l,m+1,n)+G_0(l,m-1,n)+$$
$$G_0(l,m,n+1)+G_0(l,m,n-1)=-2\delta_{l0}\delta_{m0}\delta_{n0}+2EG_0(l,m,n) \qquad ;E=3 \qquad (3)$$

In some cases one may need to use the recurrence formulae (i.e. Eq. (3)) two or three times, and by the method explained above we calculate different values for $r_1, r_2, r_3$ for $(l,m,n)$ beyond $(5,5,5)$. Various values of $\rho_1, \rho_2, \rho_3$ are shown in Table 1.

The value of the LGF at the origin (i.e. $G_o(0,0,0)$) was first evaluated by Watson in his famous paper[26], where he found that

$$G_o(0,0,0)=(\frac{2}{\pi})^2(18+12\sqrt{2}-10\sqrt{3}-7\sqrt{6})[K(k_o)]^2=0.505462.$$

where $k_o=(2-\sqrt{3})(\sqrt{3}-\sqrt{2})$

$K(k)=\int_0^{\frac{\pi}{2}} d\theta \frac{1}{\sqrt{1-k^2 Sin^2\theta}}$ is the complete elliptic integral of the first kind.

A similar result was obtained by Glasser and Zucker[27] in terms of gamma function.

To study the asymptotic behavior of the resistance in a SC, one can show that[8] as any of $l,m,n$ goes to infinity then, $G_o(l,m,n)\to 0$. Thus

$$\frac{R_o(l,m,n)}{R}\to G_o(0,0,0). \qquad (4)$$

## 3. Perturbed SC Lattice

In this section we calculate the resistance between any two lattice sites in a SC, when one of the resistors (i.e. bonds) between the sites $i_o=(i_{ox},i_{oy},i_{oz})$ and $j_o=(j_{ox},j_{oy},j_{oz})$ is broken, where[9]

$$R(i,j)=R_o(i,j)+\frac{[R_o(i,j_o)+R_o(j,i_o)-R_o(i,i_o)-R_o(j,j_o)]^2}{4[R-R_o(i_o,j_o)]} \qquad (5)$$

As an example; let us assume that the bond between $i_o = (0,0,0)$ and $j_o = (1,0,0)$ is broken. So, we calculate the resistance between any two sites. Our results are arranged in Table 2, and for example:
The resistance between the sites $i = (0,0,0)$ and $j = (1,0,0)$ is

$$R(1,0,0) = \frac{R}{2}. \tag{6}$$

i.e. the resistance between the two ends of the broken bond is $\frac{R}{2}$, which is a predictable result[9].
Now, if the broken bond is shifted to be between the sites $i = (1,0,0)$ and $j = (2,0,0)$, then one can find the resistance between any two sites (i.e. $i = (i_x, i_y, i_z)$ and $j = (j_x, j_y, j_z)$). Using Eq. (5) again one obtains the results arranged in Table 3.
For large values of $i$ *and* $j$ the resistance in a perturbed SC, becomes

$$\frac{R(i,j)}{R} \to \frac{R_o(i,j)}{R} = g_o. \tag{7}$$

We conclude that for large separation between the two sites the perturbed resistance approaches the perfect one (i.e. it approaches a finite value).

## 4. Results and Discussion

Fig. 1. shows the resistance against the site $(l,m,n)$ along the [100] direction for both a perfect infinite and perturbed SC (i.e the bond between $i_0 = (0,0,0)$ and $j_0 = (1,0,0)$ is broken). It is seen from the figure that the resistance is symmetric (i.e. $R_o(l,0,0) = R_o(-l,0,0)$) for the perfect case due to the inversion symmetry of the lattice while for the perturbed case the symmetry is broken so, the resistance is not symmetric. As $(l,m,n)$ goes away from the origin the resistance approaches its finite value for both cases[8].

Fig. 2. shows the resistance against the site $(l,m,n)$ along the [010] direction for a perfect infinite and perturbed SC (i.e the bond between $i_0 = (0,0,0)$ and $j_0 = (1,0,0)$ is broken). The figure shows that the resistance is symmetric for the perfect and perturbed case, since there is no broken

bond along this direction. As $(l,m,n)$ goes away from the origin the resistance approaches its finite value for both cases[8].

Fig. 3. shows the resistance against the site $(l,m,n)$ along the [111] direction for a perfect and for a perturbed SC (i.e the bond between $i_0 = (0,0,0)$ and $j_0 = (1,0,0)$ is broken). The resistance is symmetric along $[111]$ direction for both the perfect and perturbed cases.

Figures. 4-6. same as the above figures except that the broken bond is shifted (i.e the bond between $i_0 = (1,0,0)$ and $j_0 = (2,0,0)$ is broken). The resistance along $[100]$ direction is not symmetric in the perturbed case since the broken bond is taken to be along that direction.

From Figs. 1-6, as the broken bond is shifted from the origin along [100] direction then the resistance of the perturbed SC approaches that of the perfect lattice. Also, one can see that the perturbed resistance is always larger than the perfect one. Measurement of the resistance of a finite SC is under investigation in order to compare results.

## Figure Captions

**Fig. 1 The resistance on the perfect (squares) and the perturbed (circles) SC between $i = (0,0,0)$ and $j = (j_x, 0,0)$ along the [100] direction as a function of $j_x$. The ends of the removed bond are $i_o = (0,0,0)$ and $j_o = (1,0,0)$.**

**Fig. 2 The resistance on the perfect (squares) and the perturbed (circles) SC between $i = (0,0,0)$ and $j = (0, j_y, 0)$ along the [010] direction as a function of $j_y$. The ends of the removed bond are $i_o = (0,0,0)$ and $j_o = (1,0,0)$.**

**Fig. 3 The resistance on the perfect (squares) and the perturbed (circles) SC between $i = (0,0,0)$ and $j = (j_x, j_y, j_z)$ along the [111] direction as a function of $j$. The ends of the removed bond are $i_o = (0,0,0)$ and $j_o = (1,0,0)$.**

**Fig. 4 The resistance on the perfect (squares) and the perturbed (circles) SC between $i = (0,0,0)$ and $j = (j_x, 0,0)$ along the [100] direction as a function of $j_x$. The ends of the removed bond are $i_o = (1,0,0)$ and $i_o = (2,0,0)$.**

**Fig. 5 The resistance on the perfect (squares) and the perturbed (circles) SC between $i = (0,0,0)$ and $j = (0, j_y, 0)$ along the [010] direction as a function of $j_y$. The ends of the removed bond are $i_o = (1,0,0)$ and $i_o = (2,0,0)$.**

**Fig.3.6 The resistance on the perfect (squares) and the perturbed (circles) SC between $i = (0,0,0)$ and $j = (j_x, j_y, j_z)$ along the [111] direction as a function of $j$. The ends of the removed bond are $i_o = (1,0,0)$ and $i_o = (2,0,0)$.**

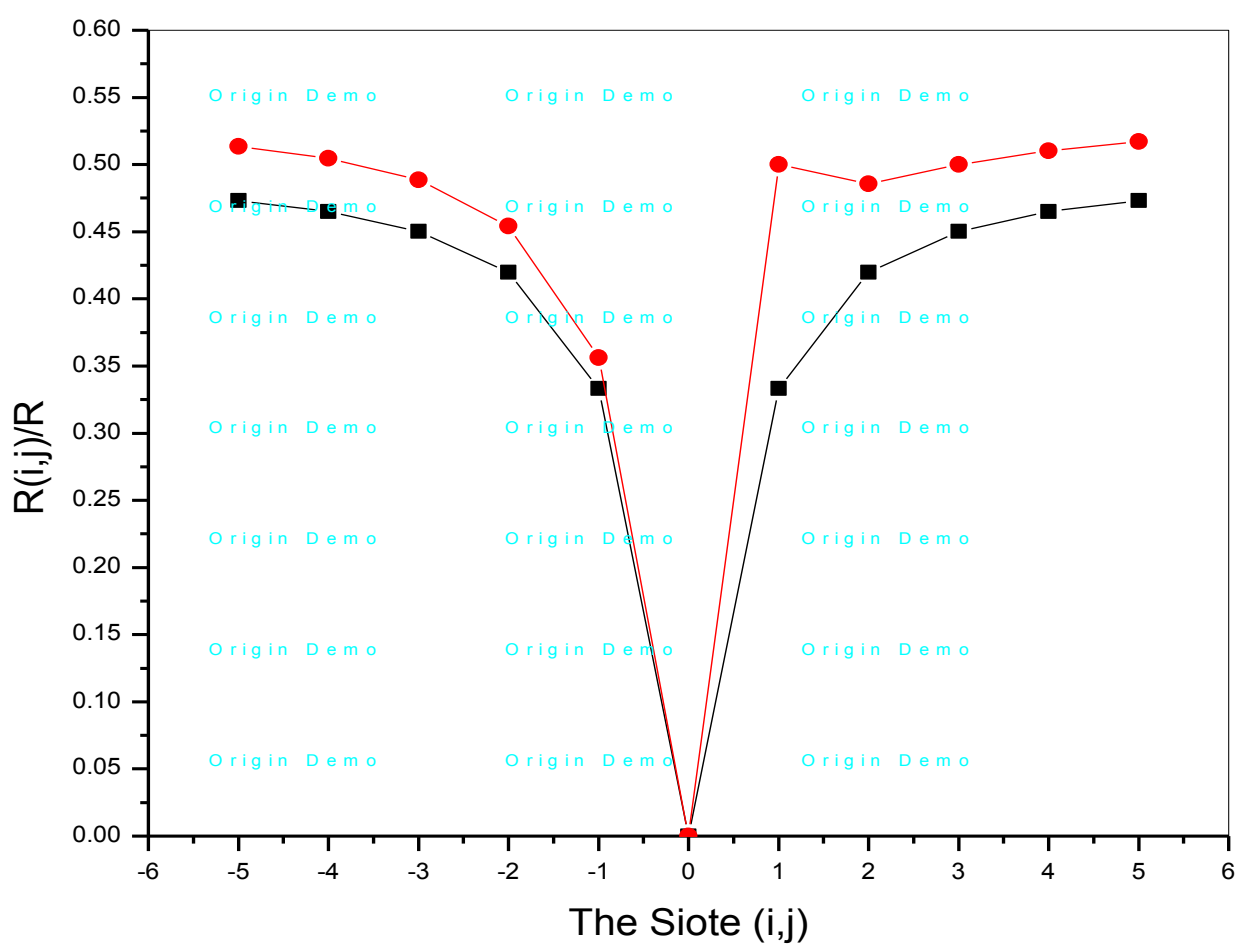

Fig. 1

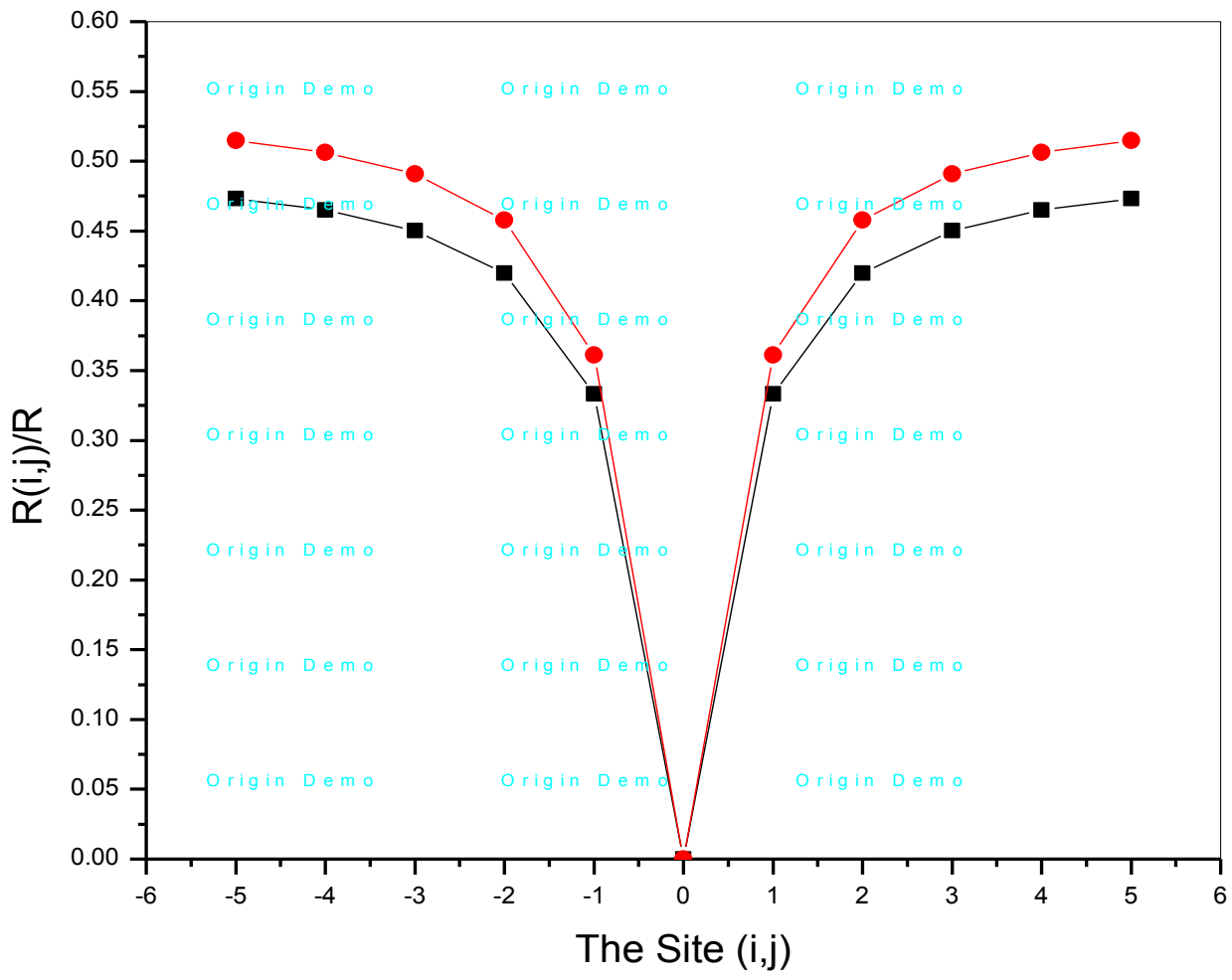

Fig. 2

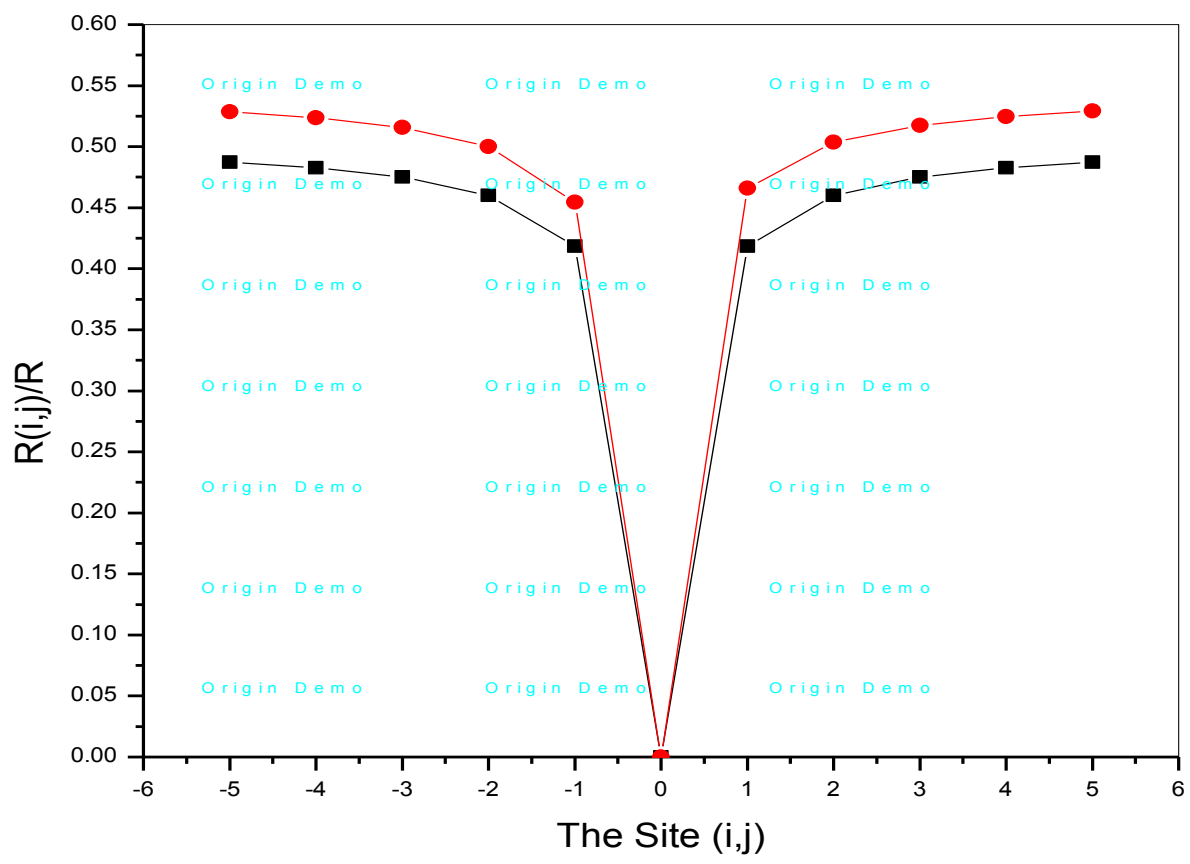

Fig. 3

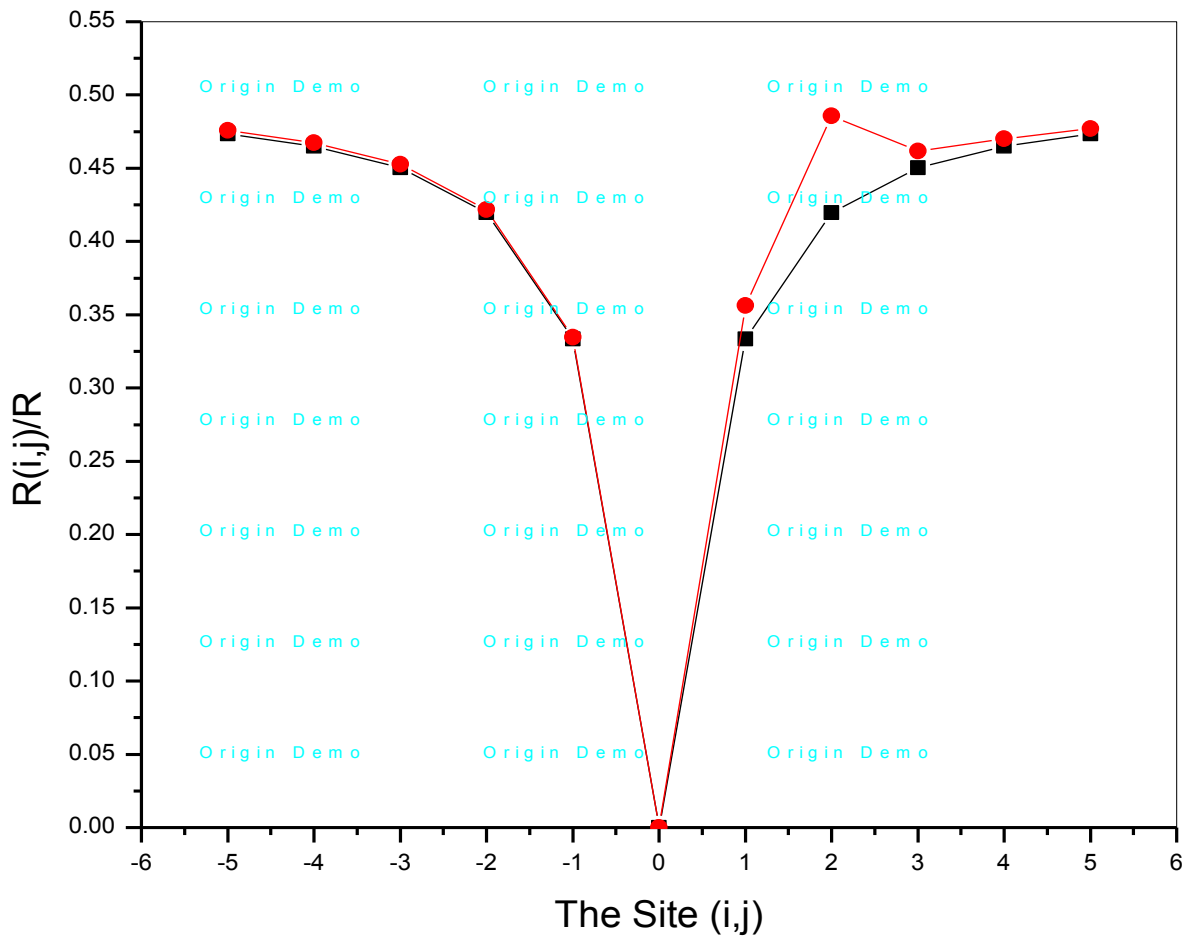
R(i,j)/R
The Site (i,j)
0.55
0.50
0.45
0.40
0.35
0.30
0.25
0.20
0.15
0.10
0.05
0.00
-6
-5
-4
-3
-2
-1
0
1
2
3
4
5
6

Fig. 4

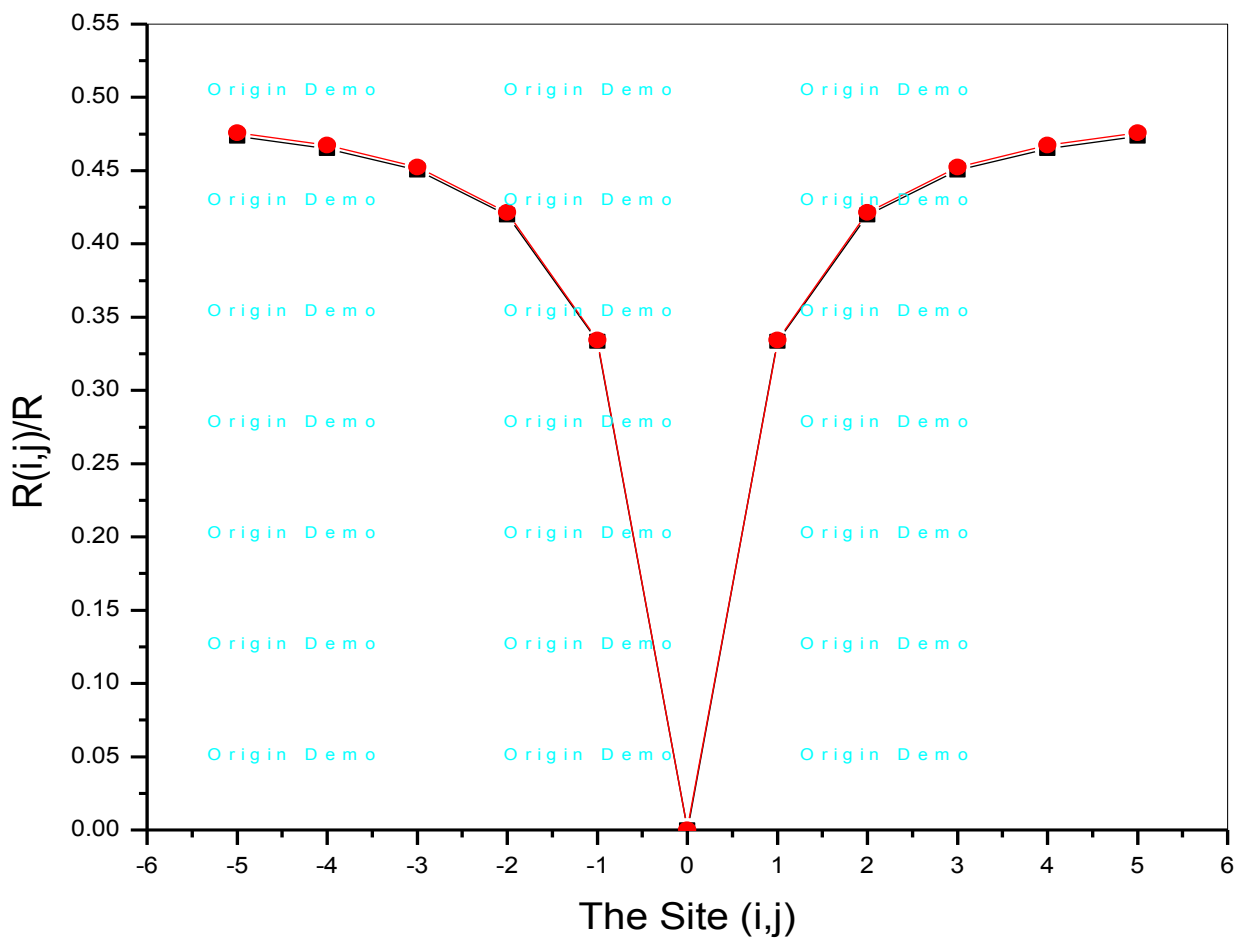
R(i,j)/R
The Site (i,j)
0.55
0.50
0.45
0.40
0.35
0.30
0.25
0.20
0.15
0.10
0.05
0.00
-6
-5
-4
-3
-2
-1
0
1
2
3
4
5
6

Fig. 5

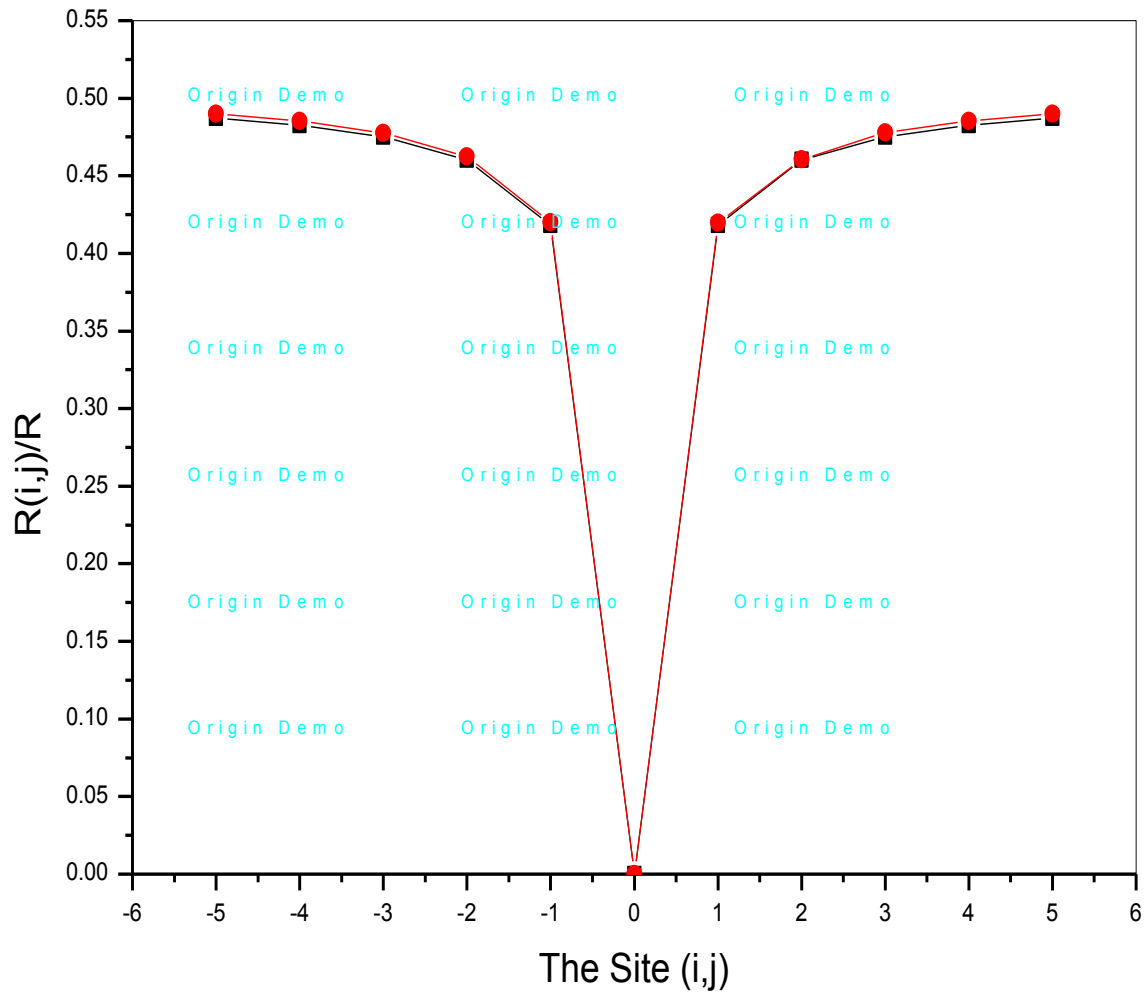

Fig. 6

# Table Captions

**Table 1: Various values of the resistance in a perfect infinite SC for arbitrary sites.**

**Table 2: Calculated values of the resistance between the sites $i=(0,0,0)$ and $j=(j_x,j_y,j_z)$, for a perturbed SC ( the bond between $i_0=(0,0,0)$ and $j_o=(1,0,0)$ is broken).**

**Table 3: Calculated values of the resistance between the sites $i=(0,0,0)$ and $j=(j_x,j_y,j_z)$, for a perturbed SC ( the bond between $i_0=(1,0,0)$ and $j_o=(2,0,0)$ is broken).**

**Table 1**

| $lmn$ | $\rho_1$ | $\rho_2$ | $\rho_3$ | $\frac{R_0(l,m,n)}{R}=\rho_1 g_0+\frac{\rho_2}{\pi^2 g_0}+\rho_3$ |
|---|---|---|---|---|
| 000 | 0 | 0 | 0 | 0 |
| 100 | 0 | 0 | 1/3 | 0.333333 |
| 110 | 7/12 | 1/2 | 0 | 0.395079 |
| 111 | 9/8 | -3/4 | 0 | 0.418305 |
| 200 | -7/3 | -2 | 2 | 0.419683 |
| 210 | 5/8 | 9/4 | -1/3 | 0.433598 |
| 211 | 5/3 | -2 | 0 | 0.441531 |
| 220 | -37/36 | 29/6 | 0 | 0.449351 |
| 221 | 31/16 | -21/8 | 0 | 0.453144 |
| 222 | 3/8 | 27/20 | 0 | 0.460159 |
| 300 | -33/2 | -21 | 13 | 0.450371 |
| 310 | 115/36 | 85/6 | -4 | 0.454415 |
| 311 | 15/4 | -21/2 | 2/3 | 0.457396 |
| 320 | -271/48 | 119/8 | 1/3 | 0.461311 |
| 321 | 161/36 | -269/30 | 0 | 0.463146 |
| 322 | -19/16 | 213/40 | 0 | 0.467174 |
| 330 | -47/3 | 1046/25 | 0 | 0.468033 |
| 331 | 38/3 | -148/5 | 0 | 0.469121 |
| 332 | -26/9 | 1012/105 | 0 | 0.471757 |
| 333 | 51/16 | -1587/280 | 0 | 0.475023 |
| 400 | -985/9 | -542/3 | 92 | 0.464885 |
| 410 | 531/16 | 879/8 | -115/3 | 0.466418 |
| 411 | 11/2 | -357/5 | 12 | 0.467723 |
| 420 | -2111/72 | 13903/300 | 6 | 0.469777 |
| 421 | 245/16 | -1251/40 | -1 | 0.470731 |
| 422 | -32/3 | 1024/35 | 0 | 0.473076 |
| 430 | -2593/48 | 28049/200 | -1/3 | 0.473666 |
| 431 | 1541/36 | -110851/1050 | 0 | 0.474321 |
| 432 | -493/32 | 4617/112 | 0 | 0.476027 |
| 433 | 667/72 | -8809/420 | 0 | 0.478288 |
| 440 | -5989/36 | 620161/1470 | 0 | 0.477378 |
| 441 | 4197/32 | -919353/2800 | 0 | 0.477814 |
| 442 | -2927/48 | 31231/200 | 0 | 0.479027 |
| 443 | 571/32 | -119271/2800 | 0 | 0.480700 |
| 444 | -69/8 | 186003/7700 | 0 | 0.482570 |
| 500 | -9275/12 | -3005/2 | 2077/3 | 0.473263 |
| 510 | 11653/36 | 138331/150 | -348 | 0.473986 |
| 511 | -271/4 | -5751/10 | 150 | 0.474646 |
| 520 | -2881/16 | 15123/200 | 229/3 | 0.475807 |
| 521 | 949/12 | -27059/350 | -24 | 0.476341 |
| 522 | -501/8 | 4209/28 | 2 | 0.477766 |
| 530 | -3571/18 | 1993883/3675 | -8 | 0.478166 |
| 531 | 1337/8 | -297981/700 | 4/3 | 0.478565 |
| 532 | -2519/36 | 187777/1050 | 0 | 0.479693 |
| 533 | 2281/48 | -164399/1400 | 0 | 0.481253 |
| 540 | -18439/32 | 28493109/19600 | 1/3 | 0.480653 |
| 541 | 1393/3 | -286274/245 | 0 | 0.480920 |
| 542 | -7745/32 | 1715589/2800 | 0 | 0.481798 |
| 543 | 5693/72 | -4550057/23100 | 0 | 0.483012 |
| 544 | -1123/32 | 560001/6160 | 0 | 0.484441 |
| 550 | -196937/108 | 101441689/22050 | 0 | 0.483050 |
| 551 | 12031/8 | -18569853/4900 | 0 | 0.483146 |
| 552 | -1681/2 | 5718309/2695 | 0 | 0.483878 |
| 553 | 5175/16 | -2504541/3080 | 0 | 0.484777 |
| 554 | -24251/312 | -1527851/7700 | 0 | 0.485921 |
| 555 | 9459/208 | -12099711/107800 | 0 | 0.487123 |
| 600 | -34937/6 | -313079/25 | 5454 | 0.478749 |
| 610 | 71939/24 | 160009/20 | -9355/3 | 0.479137 |
| 633 | 18552/72 | -747654/1155 | 0 | 0.483209 |
| 644 | -388051/1872 | 23950043/46200 | 0 | 0.486209 |
| 655 | 13157/78 | -5698667/13475 | 0 | 0.488325 |
| 700 | -553847/12 | 5281913/50 | 44505 | 0.482685 |

**Table 2**

| $j=(j_x,j_y,j_z)$ | $R(i,j)/R$ | $j=(j_x,j_y,j_z)$ | $R(i,j)/R$ |
|---|---|---|---|
| (1,0,0) | 0.5 | (-1,0,0) | 0.356208 |
| (2,0,0) | 0.485733 | (-2,0,0) | 0.454031 |
| (3,0,0) | 0.500062 | (-3,0,0) | 0.4526508 |
| (4,0,0) | 0.510257 | (-4,0,0) | 0.467337 |
| (5,0,0) | 0.5170241 | (0,-1,0) | 0.360993 |
| (0,1,0) | 0.360993 | (0,-2,0) | 0.457943 |
| (0,2,0) | 0.457943 | (0,-3,0) | 0.491033 |
| (0,3,0) | 0.491033 | (0,-4,0) | 0.506167 |
| (0,4,0) | 0.506167 | (0,-5,0) | 0.5147228 |
| (0,5,0) | 0.5147228 | (0,0,-1) | 0.360993 |
| (0,0,1) | 0.360993 | (0,0,-2) | 0.457943 |
| (0,0,2) | 0.457943 | (0,0,-3) | 0.491033 |
| (0,0,3) | 0.491033 | (0,0,-4) | 0.506167 |
| (0,0,4) | 0.506167 | (0,0,-5) | 0.5147228 |
| (0,0,5) | 0.5147228 | (-1,-1,-1) | 0.454367 |
| (1,1,1) | 0.4659804 | (-2,-2,-2) | 0.50009 |
| (2,2,2) | 0.503597 | (-3,-3,-3) | 0.5158855 |
| (3,3,3) | 0.517510166 | (-4,-4,-4) | 0.5237707 |
| (4,4,4) | 0.524705 | (-5,-5,-5) | 0.528490451 |
| (-5,0,0) | 0.513569 | (5,5,5) | 0.5290906 |

**Table 3**

| $j=(j_x,j_y,j_z)$ | $R(i,j)/R$ | $j=(j_x,j_y,j_z)$ | $R(i,j)/R$ |
|---|---|---|---|
| (1,0,0) | 0.356208 | (-1,0,0) | 0.334495 |
| (2,0,0) | 0.485733 | (-2,0,0) | 0.421618 |
| (3,0,0) | 0.461555 | (-3,0,0) | 0.452650 |
| (4,0,0) | 0.470021 | (-4,0,0) | 0.467337 |
| (5,0,0) | 0.477057 | (-5,0,0) | 0.475810 |
| (0,1,0) | 0.334191 | (0,-1,0) | 0.334191 |
| (0,2,0) | 0.421552 | (0,-2,0) | 0.421552 |
| (0,3,0) | 0.452738 | (0,-3,0) | 0.452738 |